\documentclass[aps,prb,twocolumn,showpacs,longbibliography]{revtex4-1}

\usepackage{color}

\usepackage[normalem]{ulem}
\newcommand{\stkout}[1]{\ifmmode\text{\sout{\ensuremath{#1}}}\else\sout{#1}\fi}

\usepackage{graphicx}

\begin{document}
\newcommand{\ud}{{\mathrm d}}
\newcommand{\sech}{\mathrm{sech}}

\title{Temperature gradient and thermal conductivity in superdiffusive materials}

\author{Yuanyang Ren}
\affiliation{State Key Lab. of Electrical Insulation and Power Equipment, Xi'an Jiaotong University, No. 28 Xianning West Road, Xi'an 710049, Shaanxi, China}

\author{Kai Wu}
\affiliation{State Key Lab. of Electrical Insulation and Power Equipment, Xi'an Jiaotong University, No. 28 Xianning West Road, Xi'an 710049, Shaanxi, China}

\author{David Cubero}
\email[]{dcubero@us.es}
\affiliation{Departamento de F\'{\i}sica Aplicada I, EUP, Universidad de Sevilla, Calle Virgen de \'Africa 7, 41011 Sevilla, Spain}


\begin{abstract}
Thermal conductivities are routinely calculated in molecular dynamics simulations by keeping the boundaries at different temperatures and measuring the slope of the temperature profile in the bulk of the material,  explicitly using Fourier's law of heat conduction.  Substantiated by the observation of a distinct linear profile at the center of the material, this approach has also been frequently used in superdiffusive materials, such as nanotubes or polymer chains, which do not satisfy Fourier's law  at the system sizes considered. It has been recently argued that this temperature gradient procedure yields worse results when compared with a method based on the temperature difference at the boundaries---thus taking into account the regions near the boundaries where the temperature profile is not linear. We study a realistic example, nanocomposites formed by adding boron nitride nanotubes to a polymer matrix of amorphous polyethylene, to show 
that in superdiffusive materials, despite the appearance of a central region with a linear profile, the temperature gradient method is actually inconsistent with a conductivity that depends on the system size, and, thus, it should be only used in normal diffusive systems.  
\end{abstract}

\maketitle

\section{Introduction}

Fourier's law is only guaranteed for materials where heat transport is diffusive. In a pure harmonic systems, it is well known \cite{lebo67} that transport is ballistic, and thus the thermal conductivity diverges as the system size $L\rightarrow\infty$. It came as a big surprise when it was shown, in 1955 by Fermi, Pasta and Ulam in a celebrated numerical study\cite{fpu55},  that one-dimensional (1D) anharmonic lattices also exhibits an infinite thermal conductivity, for divergent behavior was only expected from harmonic lattices, where phonon scattering is absent. The Fermi-Pasta-Ulam lattice was the first example \cite{dhar08,lepri20}, of a nonlinear lattice displaying {\em superdiffusive} behavior, defined as a thermal conductivity $\kappa$ that diverges as $\kappa\sim L^\beta$, where $0<\beta<1$ ($\beta=0$ corresponds to normal diffusive materials satisfying Fourier's law and $\beta=1$ to ballistic transport).

Generally, superdiffusion is a kind of anomalous diffusion \cite{klages08} 
 in which the object studied, such as the probability of the position of a material particle or the energy in a group of particles in our present case, spread in space in average much faster than the rate given by normal diffusion---which, in the absence of external agents, is precisely described by a parabolic partial differential equation, called Fick or Smoluchowski equation in the context of particle motion, and heat equation in heat conduction. Similarly, subdiffusion is characterized by a slower spread. Quantitatively, anomalous diffusion for particle motion is defined from the exponent $\alpha$ associated to the divergence with time, for $t\rightarrow\infty$, of the mean square displacement $\langle x^2(t)\rangle-\langle x(t)\rangle^2\sim t^\alpha$, requiring $\alpha<1$ for subdiffusion and $\alpha>1$ for superdiffusion, whereas for heat transport it is usually defined from the conductivity's exponent $\beta$. Both are connected via the spread of the mean square deviation of energy  $\langle x^2(t)\rangle_E-\langle x(t)\rangle_E^2\sim t ^\alpha$,  where the averages are taken with respect an excess energy perturbation distribution\cite{hanbao14}, and $\alpha=1+\beta$.
 Examples are numerous \cite{klages08,cub12,fin20,lepri20,dhar08}.

Low dimensional structures, such as nanowires, carbon nanotubes (CNTs) or boron nitride nanotubes (BNNTs), can be regarded as real-world manifestations of heat superdiffusive materials, with numerous experiments demonstrating superdiffusive behavior \cite{chang08,bala10,meimen14,donba14,bouess16,chang17}. 
We focus here on quasi-1D materials---in two-dimensional lattices the divergence, instead of a power law, is known to be logarithmic \cite{lepri20}, $\kappa\sim\log(L)$, which has been corroborated experimentally with graphene sheets\cite{donba14}. 
There is still some controversy \cite{qinkoj17,liuli17,lepri20,saas15,upaaks16,ala19,brunoj20} of whether the anomalous behavior stands up in these materials for an arbitrarily large system size $L$, or rather the thermal conductivity saturates to a constant value once $L$ is large enough---indicating eventual normal diffusion.

Nevertheless, we take a practical approach here, and regard as heat superdiffusive behavior whenever the material exhibits a monotonically-increasing, system-size dependence in the thermal conductivity $\kappa(L)$, as observed in these nano-materials with sizes, at least, up to a few microns.


While 
a certain sensitivity to the nature of its surroundings is likely to be present in a superdiffusive material, it is desirable that this dependence be small for a meaningful, useful definition of the thermal conductivity.

On the other hand, BNNTs have the advantage over some CNTs, for our purposes here, that the thermal conductivity is exclusively due to phonons \cite{chang08}, whereas in CNTs there can be a small, but non-negligible contribution due to electrons, thus complicating the theoretical analysis.

Nonequlibrium Molecular Dynamics (NEMD) simulations have been frequently used  to study these nano systems\cite{maru02,baowen05,saas15,zhang20}. The common procedure of thermostatting the boundaries, waiting for a stationary profile, and then computing the thermal conductivity from Fourier's law, applied at a central region of the material, where a linear temperature profile is observed, away from the boundaries, has been applied repeatedly in the past \cite{liu12,maru02,baowen05,donba14,liu18,luo18,zhang20}. This method, which we refer to as the {\em temperature-gradient} method, has been a standard procedure to minimize the effect of the boundaries in the calculation.

However, it has been very recently pointed out \cite{dona19} that this procedure should not be used, and the thermal conductivity should be computed instead from the temperature difference of the boundaries' baths, thus including regions closer to the boundaries, where the temperature profile is non-linear. The proposed method is supported by a comparison of the NEMD results, of carbon-based nano structures, with results using an atomistic Green function method in the ballistic regime and homogeneous nonequilibrium simulations. The comparison outside the ballistic regime is based on a phenomenological theory  \cite{dona19,ala19,saas15}, favoring the temperature-difference method. Here we  prove numerically within the framework of NEMD simulations, and thus without relying on other approximate methods, that the temperature-gradient method is actually inconsistent when the thermal conductivity depends on the system size.

In this work, we study BNNTs immersed in a polymer matrix based on amorphous polyethylene (PE).
 Heat transport in the polymeric material is of the normal diffusive type, in fact it has a low thermal conductivity \cite{rajmut18} $\kappa_\mathrm{PE}=$0.5 W/mK---for this reason it is frequently referred as a thermal insulator. Composites are created with additional coupling agents attached covalently to both the nanotube and the polymer matrix with various densities. The coupling agents slow down transport in the nanotube, thus inducing a transition in the composite from superdiffusive to diffusive behavior. This setup allows us to highlight the different physical explanations yielded by the temperature- gradient and temperature-difference methods. We then show that, unlike the temperature-difference procedure, the temperature-gradient method is actually inconsistent with the observed phenomena.

The paper is organized as follows. In Sec.~\ref{sec:model} we introduce the model and simulation techniques. In Sec.~\ref{sec:tempmethods} we discuss the temperature-gradient and temperature-different methods, and the different physical pictures of heat transport associated with them in our realistic example. The temperature-gradient method is shown to be inconsistent in Sec.~\ref{sec:incon}, where we also briefly discuss the implications of a theory based on phonon transport.
Finally, Sec.~\ref{sec:con} ends with the main conclusions.


\section{Model and methods}
\label{sec:model}

NEMD simulations were carried out using the software LAMMPS \cite{26}.  
In order to accurately describe the heat transport properties of BNNTs and PE, we have used the Tersoff \cite{27,31} potential to model the bond interaction between boron (B) and nitrogen (N), and the OPLS-AA force field \cite{28} to describe the covalent bond and non-bonding interactions for polyethylene. The polymer matrix was modeled using 60 n-alkane chains, each one with 500 carbons. 

BNNTs are made of hexagonal networks of BN bonds. We have considered single walled BNNTs with an armchair (10,10) BN structure ---a common BNNT with a diameter $d=1.36$~nm. Interlayer interactions between BNNTs and PE were modeled using Lennard-Jones potentials \cite{uff92}  
with a parametrization based on the geometric mixing rules.

\begin{figure}[h]
 \centering
 \includegraphics[width=9cm]{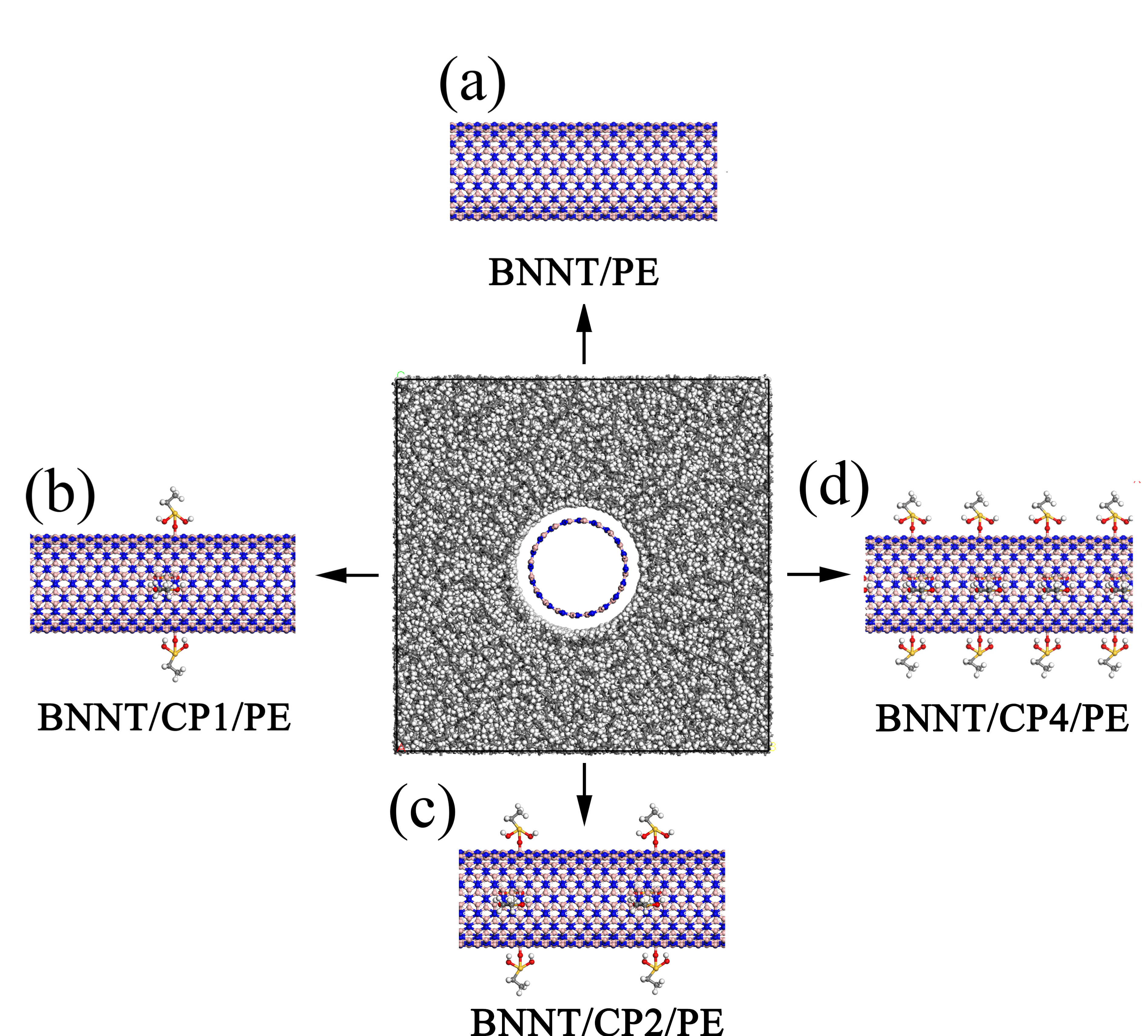}
 \caption{
Schematic view of nanocomposite structures. PE atoms are only shown in the central figure. 
(a) BNNT/PE without agents. (b) With one couping agent per nanometer of BNNT (named as BNNT/CP1/PE) . (c) With double amount of  coupling agents (BNNT/CP2/PE). (d)  With four times the amount of coupling agents (BNNT/CP4/PE). 
}
 \label{fig:1}
\end{figure}

The silane coupling agent modeled in the simulations was based on vinyltrimethoxysilane, which has the ability to form strong bonds both with BNNT \cite{18} and PE \cite{mahajan01,32}. We have considered that each silane agent is bonded to a single atom B at the nanotube, as shown in Fig.~\ref{fig:1}, and with 90\% of them being also bonded to PE chains, replacing the alkylene group present in the isolated form of the agent with a PE chain. A combination of a high temperature (500~K) NVT simulation and a subsequent NPT simulation ---keeping the atoms at the BNNT frozen--- was carried out to melt the polymer matrix and generate amorphous PE at the experimental density 0.86~g/cm$^3$. The nanocomposites studied in this paper are illustrated in Fig.~\ref{fig:1}.

Periodic boundary conditions were applied in all three directions. In all setups, the atomic coordinates were equilibrated using a NPT simulation, at 300~K and 1~atm (with damping time 0.5~ps), for 0.5~ns, and then  further relaxed in the NVT ensemble (300 K) for another 0.5~ns. Then, a steady heat flux was imposed among the symmetry axis of the BNNT by thermostatting two regions with several layers of atoms (both of BNNT and PE in the composites), each region with thickness 10 nm, to the temperatures $T_h$ and $T_c$, using either a velocity rescaling algorithm or a Langevin bath with a damping time of $\tau=0.5$~ps. Unless explicitly stated, the thermostatted temperatures were $T_c=280$~K and $T_h=320$~K. Next to the bath regions, all atoms in the two utmost 1.5~nm of layers were fixed to avoid heat exchange across the periodic boundary in that direction.
During the first 1~ns of simulation, the temperature profile--measured locally using its equipartition definition---was always observed to reach a steady state. We then used a further 2~ns interval to sample the temperature profile and heat flux along the main axis of the BNNT, taken as the $z$-direction. The heat flux was computed by measuring the energy rate transferred to the material at the thermostatted regions, and then dividing by the cross-section area $A$. 

In the composite simulations the cross-section area is given by the simulation box, $A=L_x L_y$, whereas for simulations of bare BNNT this quantity is somehow arbitrary. In this case, following Ref.~\cite{31}, we use the expression  $A=\pi h d$ for a hollow tube of diameter $d$ with a small tube thickness $h$,  taking the thickness $h=0.335$~nm from the estimation based on the van der Waals interaction.


\section{The temperature-gradient and temperature-difference methods}
\label{sec:tempmethods}

In the presence of a steady temperature gradient, the thermal conductivity $\kappa$ is commonly  computed via Fourier's law,
\begin{equation}
j =- \kappa \frac{ \partial T}{\partial z},
\label{eq:fourier}
\end{equation}
where the heat flux $j$, time and coordinate independent at the steady state, can be easily determined from the energy exchange at the bath regions.  In normal diffusive systems, the heat diffusion equation---obtained by combining (\ref{eq:fourier}) with the energy conservation law---predicts a linear temperature profile $\partial T/\partial z=$constant, which is usually observed in the simulations in a region---the bulk---away from the boundaries. Near the boundaries the profile is rarely linear, which is often interpreted as the action of a phonon mismatch between the phonon distribution at the bulk and the boundary regions\cite{lepri03}, creating an additional thermal resistance at the boundaries, also known as Kapitza resistance. The standard procedure then, in order to minimize boundary effects,  is to use (\ref{eq:fourier}) with the observed value of the temperature gradient $\partial T/\partial z$ at the bulk, yielding
\begin{equation}
\kappa_\mathrm{TG}=-j / \left(\frac{\partial T}{\partial z}\right)_\mathrm{bulk}.
\label{eq:ktg}
\end{equation}

\begin{figure}[h]
 \centering
 \includegraphics[width=7cm,height=4cm]{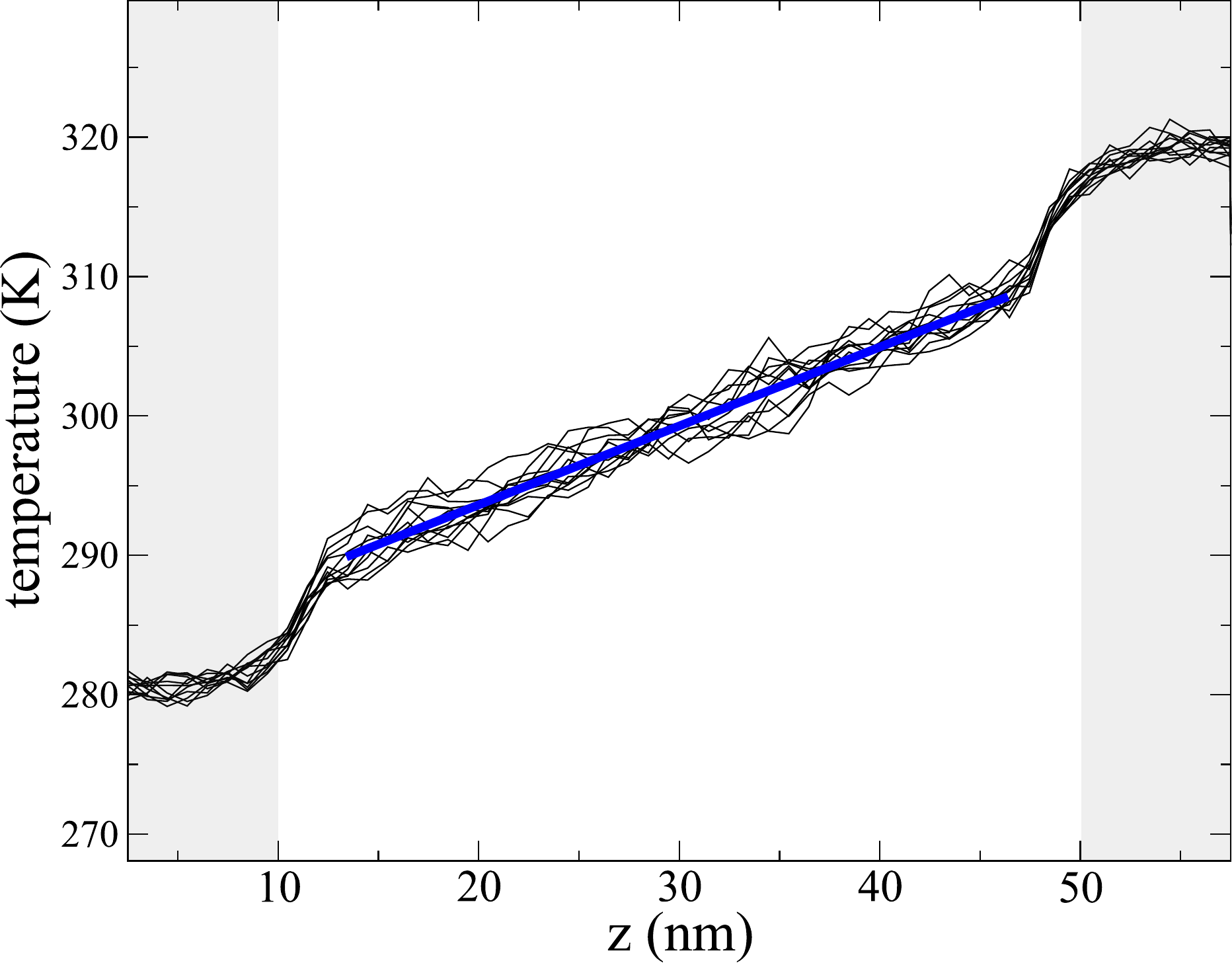}
 \caption{
Typical temperature profile. The black solid lines show the local temperature along the symmetry axis of the nanotube in a bare BNNT simulation of length $60$~nm for several instants of time. The thermostatted regions are shaded. The length of the non-thermostatted region is $40$~nm. The blue thick solid line in the center is a straight-line fit, indicating a linear temperature gradient away from the boundaries, in a region of length $\sim 35$~nm. 
}
 \label{fig:tempgrad}
\end{figure}

Being linear in the bulk, the gradient in (\ref{eq:ktg}) can be written as $(\partial T/\partial z)_\mathrm{bulk}=(T'_h-T'_c)/L'$, where $L'$ is the length of the bulk, and $T'_h$ and $T'_c$ are the temperatures at the boundaries of the bulk. However, let us notice that in normal diffusive systems, due to the Kapitza resistance, $L'$ is in general smaller than the distance $L$ between the reservoirs, and  $T'_h$ and $T'_c$ are different from the reservoirs' temperatures $T_h$ and $T_c$.

In superdiffusive materials, such as nanotubes, it is not uncommon to find a central region where the temperature profile is also linear. An example is shown in Fig.~\ref{fig:tempgrad} for bare BNNT.  We have observed the same behavior---a central region with a clear linear temperature profile when time averaged---in all the simulations reported in this paper. It is thus very tempting \cite{liu12,baowen05,liu18,luo18,zhang20} to apply the same temperature-gradient procedure (\ref{eq:ktg}) to compute the thermal conductivity as in normal diffusive systems.

However, in low dimensional systems a linear temperature profile is not expected \cite{lepri03,dhar08,lepri20}, in fact the Fermi-Pasta-Ulam lattice \cite{fpu55} exhibits a highly non-linear profile, being customary in this context  \cite{dhar08,lepri20} to define the thermal conductivity from the imposed temperature difference at the boundaries, i.e.
\begin{equation}
\kappa_\mathrm{TD}=-j L/ (T_h-T_c).
\label{eq:ktd}
\end{equation}

Let us notice the temperature difference method (\ref{eq:ktd})  neglects the Kapitza resistance at the boundaries, and thus, in a normal diffusive system it would not necessarily provide the intrinsic thermal conductivity of the material, but a value affected by undesired finite size effects. However, this problem can be avoided---or severely reduced---in a NEMD simulation  by making the thermostatted regions large enough, as we show later on, minimizing the temperature slip at the boundaries. Moreover, a large bath region is also needed in the ballistic regime in order to ensure proper thermalization \cite{saas14,dona19}, being thus an important requirement for the application of   (\ref{eq:ktd}). A reasonable criterion \cite{saas14} with a Langevin bath, with damping time $\tau$, is to make the length of the thermostatted region larger than $v_g\tau$, where $v_g$ is the average phonon group velocity of the material.

Next, we devote the remaining of the section to show that the different methods (\ref{eq:ktg}) and (\ref{eq:ktd}) used to calculate the thermal conductivity are not simply two different definitions of the thermal conductivity, in superdiffusive materials they carry different, incompatible physical rationalizations of the same phenomena. 

\begin{figure}[h]
 \centering
 \includegraphics[width=7cm,height=4cm]{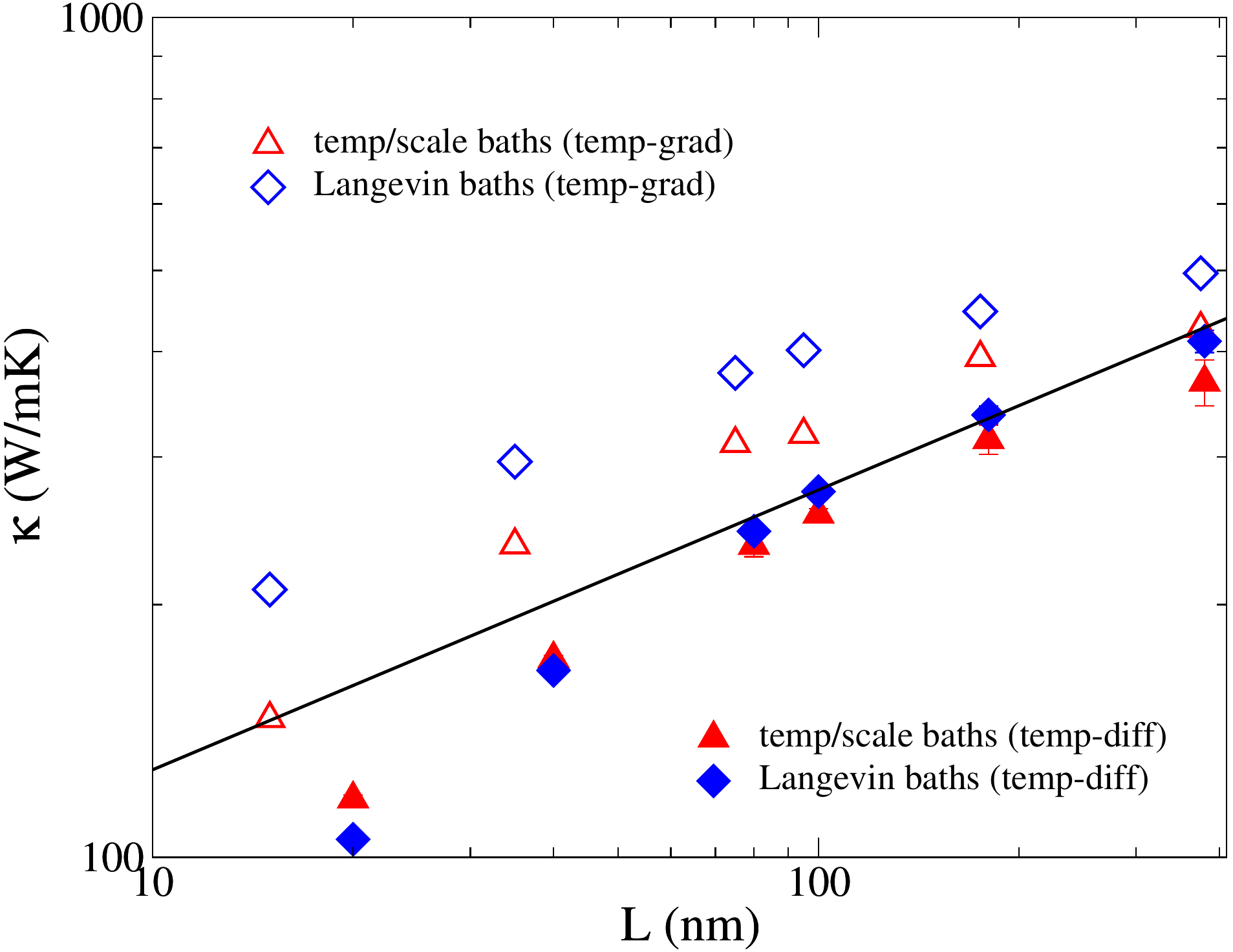}
 \caption{
Thermal conductivity of the bare BNNT as a function of its length. Empty symbols denote the temp-grad method (\ref{eq:ktg})  and filled ones the  temp-diff method (\ref{eq:ktd}).  Diamonds correspond to a Langevin bath and triangles to thermostatting via velocity rescaling. The solid line is $\kappa\sim L^\beta$,  with $\beta=1/3$, the theoretical result for a one-dimensional nonlinear system \cite{dhar08}, plotted as a reference. 
}
 \label{fig:kofL}
\end{figure}

Figure~\ref{fig:kofL} reports the conductivity obtained by the two methods applied to a single nanotube with various lengths. For the  temp-diff method (\ref{eq:ktd}), $L$ corresponds to the full length of the non-thermostatted region in the tube. For the temp-grad method (\ref{eq:ktg}). $L$ is taken as the length of the region where the profile is linear, as shown in Fig.~\ref{fig:tempgrad}. Both methods show superdiffusive behavior, but with different values and growths. 

For each method, two sets of results are shown in Fig.~\ref{fig:kofL}, each one corresponding to one way of controlling the temperature at the thermostatted regions: simply rescaling the atoms velocities (triangles) and a Langevin bath (diamonds)---in which frictional and random forces are added to the atoms. Unlike the velocity rescaling method, which is deterministic, the Langevin heat bath is stochastic and thus expected to produce better results, without undesired artifacts \cite{baowen10,dona19}. Figure~\ref{fig:kofL} shows that both bath types produce equivalent data under the temp-diff method, while with the temp-grad method the Langevin bath yields larger conductivities than the velocity rescaling one. 

The observed independence on the thermostatting technique of the temp-diff method is a first indication of its superiority over the temp-grad procedure, but it is not conclusive by itself because a certain sensitivity to poor heat baths, such as the deterministic velocity scaling method, would not be surprising in a superdiffusive material.

Figure~\ref{fig:kofL} also shows the theoretical prediction $\kappa\sim L^\beta$,  with $\beta=1/3$,  for systems which are strictly 1D\cite{dhar08,lepri20}. 
The Fermi-Pasta-Ulam lattice, among a large class of 1D models with three conservation laws\cite{dhar08,lepri20}, has been shown to follow this power law in the asymptotic limit $L\rightarrow\infty$ ---other 1D models with a symmetric anharmonic coupling fit into a different universality class\cite{lepri20} with a still unknown exponent, either $\beta=1/2$ (from hydrodynamic mode-coupling theory) or $\beta=2/5$ (kinetic theory). The results for BNNT under the temp-diff method shows in Fig.~\ref{fig:kofL} a tendency towards the 1D prediction with $\beta=1/3$, despite being strictly a three-dimensional system. However, it is not clear if this behavior will hold for larger tube lengths. In fact, it remains unclear  \cite{dongal07,henry08,saas15,ala19,brunoj20} if the thermal conductivity, associated to the common semiempirical ---many body--- potentials used to model the atomic interactions in nano-tubes, tends to a finite value or diverges with $L$, since current computer limitations restrict the simulations to a few microns at the most.

In order to further illustrate the disparities between the results of the temp-grad and temp-diff methods, let us now turn our attention to the spectral thermal conductivity, which is obtained from the spectral heat current \cite{saas15}
\begin{equation}
q(\omega)=\sum_{i\in \hat{L}}\sum_{j\in \hat{R}} q_{i\rightarrow j}(\omega), 
\end{equation}
where the sums are over two disjoint atom sets $\hat{L}$ and $\hat{R}$, partitioning the tube into the left and right halves, respectively. The heat current from one atom in one set to another in the complementary one is given by  \cite{saas15}
\begin{equation}
q_{i\rightarrow j}(\omega)=-\frac{2}{\Delta t\,\omega}\sum_{\alpha,\beta\in\{x,y,z\}}\mathrm{Im} \left( u_\alpha^{(i)}(\omega)^* K_{\alpha\beta}^{ij} u_\beta^{(j)}(\omega)  \right), 
\label{eq:qdetail}
\end{equation}
where $\Delta t$ is the simulation sampling time, $u_\alpha^{(i)}(\omega)$ is the discrete Fourier transform of the $\alpha$-component of the velocity of atom $i$, and  $K^{ij}$ is a force matrix, resulting from approximating the interatomic force between atoms $i$ and $j$ by the linear expression
\begin{equation}
F_\alpha^{ji} = -\sum_\beta x_\beta^{(i)}  K_{\beta\alpha}^{ij}, 
\label{eq:flin}
\end{equation}
being $ x_\beta^{(i)}$ the $\beta$-component of small displacements of atom $i$.  
 The  elements $K^{ij}$ are determined from the finite-difference derivatives of the interatomic potential energy function.
The total flux along the tube is obtained by summing over all frequencies 
\begin{equation}
j=\int_0^\infty d\omega \,q(\omega)/(2\pi A).
\label{eq:jtotspec}
\end{equation}
We have checked in the bare BNNT simulations that the heat flux obtained from (\ref{eq:jtotspec}) is within 5\% of the flux obtained from the energy transfer at the boundaries, indicating that the linear approximation (\ref{eq:flin}) is valid for the heat flux calculation along the nanotube---though note the atoms velocities in (\ref{eq:qdetail}) are calculated in the simulation with the full non-linear forces.

\begin{figure}[h]
 \centering
 \includegraphics[width=7cm,height=4cm]{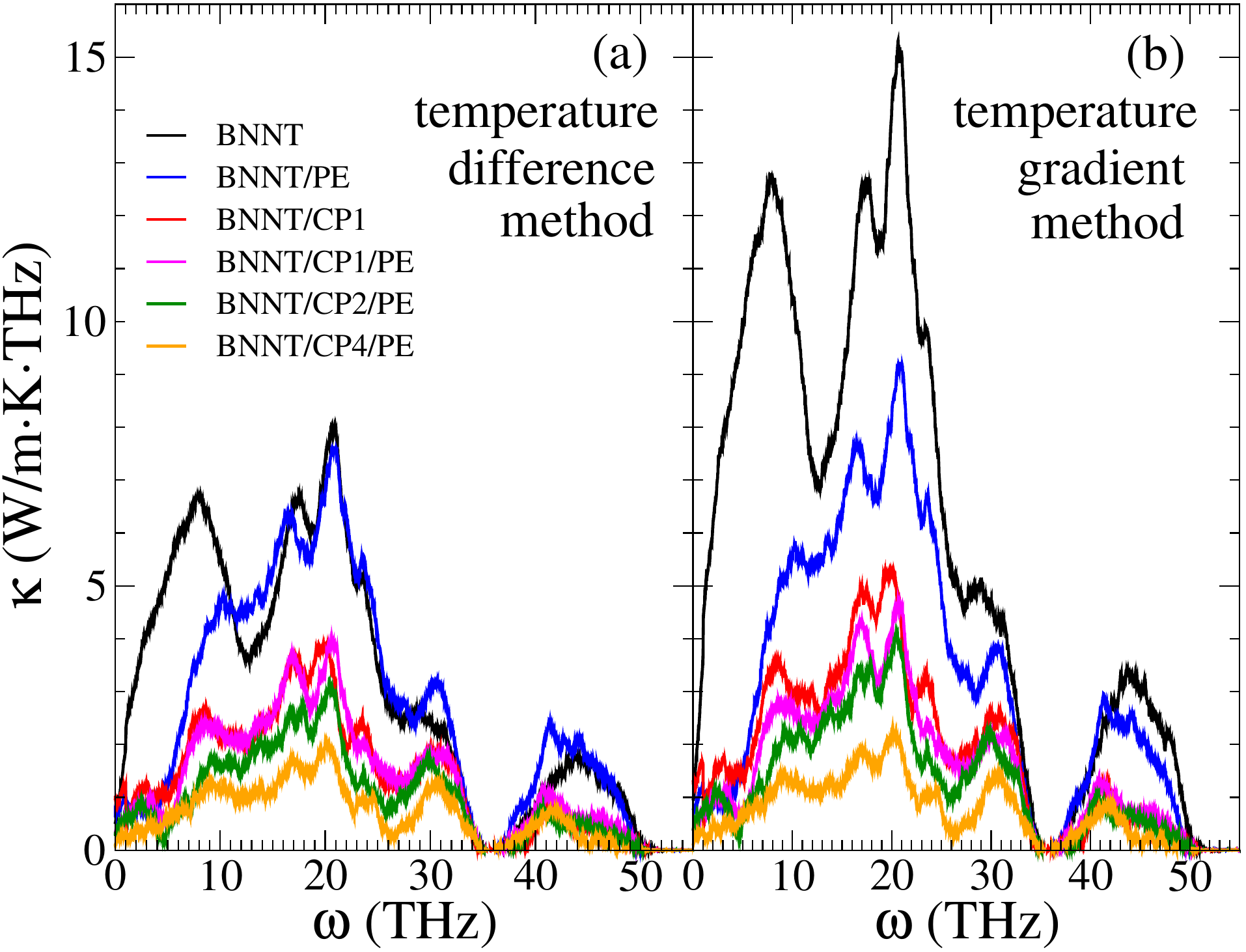}
 \caption{
Spectral thermal conductivity of BNNT in various composite systems with a non-thermostatted region of  length 40~nm. The temperatures at the thermostatted regions where set to $T_c=270$~K and $T_h=330$~K with Lagenvin heat baths. (a) temp-diff method (\ref{eq:komtd}). (b) temp-grad method (\ref{eq:komtg}). 
}
 \label{fig:kofom}
\end{figure}

The spectral thermal conductivity $\kappa(\omega)$ is naturally obtained from the spectral heat current $q(\omega)$ consistently with the prescribed procedure, i.e., for the temp-diff method we have 
 \begin{equation}
\kappa_\mathrm{TD}(\omega)=-\frac{q(\omega) L}{A (T_h-T_c)},
\label{eq:komtd}
\end{equation}
while for the temp-grad prescription, consistency requires
 \begin{equation}
\kappa_\mathrm{TG}(\omega)=-\frac{q(\omega)}{A \left(\frac{\partial T}{\partial z}\right)_\mathrm{bulk}}.
\label{eq:komtg}
\end{equation}
For each system, the difference between the spectral conductivity (\ref{eq:komtd}) and (\ref{eq:komtg}) is just a constant independent of $\omega$. Fig.~\ref{fig:kofom} shows the simulation results obtained under each procedure for the composite systems. 

Before discussing in detail the different explanations yielded by each method, let us check the consistency of the calculations, by using the results reported in Fig.~\ref{fig:kofom} to predict the thermal conductivity of the composite systems and then compare them with direct results from NEMD simulations with an smaller temperature difference $T_h-T_c$. 

The total conductivity can be obtained from the spectral conductivity by summing over all frequencies, $k=\int_0^\infty d\omega \,\kappa(\omega)/(2\pi)$. Note that in Fig.~\ref{fig:kofom} only the spectral conductivities of the nanotube are considered. Due to the var der Waals interaction with the polymer matrix, the conductivity of BNNT is calculated to be reduced---compared to the bare case---to a 88\% of its original value under the temp-diff method ( 56\% under the temp-grad method), when the nanotube is surrounded by the polymer matrix. The addition of coupling agents, attached covalently to both the BNNT and the polymer matrix, further slows down heat transport in the tube, reducing its conductivity to 45\%, 33 \%, and 22\% of the bare tube's value, corresponding to the composites BNNT/CP1/PE, BNNT/CP2/PE, and BNNT/CP4/PE, respectively, under the temp-diff method (under the temp-grad the percentages obtained are 28\%, 22\%, and 20\%, respectively).
Since the total flux in the composite is the sum of the contributions due to the nanotube and polymer matrix, we can use the results of the simulations reported in  Fig.~\ref{fig:kofom} to predict the full conductivities of the composites with the formula
\begin{equation}
k^\mathrm{pred}=k_\mathrm{BNNT}\frac{\pi h d}{L_x L_y}+k_\mathrm{PE}\frac{L_xL_y-\pi(d/2+h)^2}{L_x L_y},
\label{eq:k:pred}
\end{equation}
where $\kappa_\mathrm{BNNT}$ is the conductivity of BNNT in the composite, and $k_\mathrm{PE}$ is approximated by its value in the isolated case.
Table~\ref{tab:k:composite} shows both the thermal conductivites of the composites under the temperature difference  $T_h-T_c=40$~K, and the predictions using (\ref{eq:k:pred}) and the data obtained with $T_h-T_c=60$~K. Both methods show very good agreement between the predictions and the direct measures. This comparison also serves as a confirmation that the flux $j$ is indeed proportional to the temperature difference $T_h-T_c$, as well as the temperature gradient in the central region. 

\begin{table}[h]
\small
  \caption{
Full thermal conductivities (in W/mK) of various composite systems, all with a non-thermostatted region of length 40~nm. The temperatures at the thermostatted regions where set to $T_c=280$~K and $T_h=320$~K with a Langevin bath.}
  \label{tab:k:composite}
  \begin{tabular*}{0.48\textwidth}{@{\extracolsep{\fill}}lllll}
    \hline
     & $\kappa_\mathrm{TD}$  &  $\kappa_\mathrm{TD} ^\mathrm{pred}$   & $\kappa_\mathrm{TG}$  &  $\kappa_\mathrm{TG} ^\mathrm{pred}$  \\
    \hline
    PE                              &         0.5       &               &  0.5     \\
    BNNT                        &        167      &               &  296    \\
    BNNT/PE                  &        12        &   12.1   &  13.6 & 13.5   \\
    BNNT/CP1/PE        &         6          &   6         &   6.5   &  6.4 \\
    BNNT/CP2/PE        &       4.5        &    4.5    &  4.6    & 4.9\\
    BNNT/CP4/PE        &       3.3        &   3.2     &  3.4     & 3.3\\
    \hline
  \end{tabular*}
\end{table}

Let us now focus on the differences between the temp-diff and temp-grad methods.
In the spectral data of Fig.~\ref{fig:kofom}, both methods only show good agreement in the composite BNNT/CP4/PE, the one with the highest concentration of coupling agents attached covalently to both the (superdiffusive) BNNT and the (diffusive) polymer matrix PE, and thus with the smallest conductivity. This agreement is also confirmed in Tab.~\ref{tab:k:composite}, where the conductivities of BNNT/CP4/PE and bare PE are almost the same with both methods, a fact that can be attributed to the  considerably long extent---about 10 nm---of the thermostatted regions, which minimizes Kapitza resistance.

The largest disagreement between both method's results is found in bare BNNT, the most superdiffusive system, where the temp-grad method provides a considerably larger spectral conductivity than the temp-diff one. 

But also it can be observed a notable disparity in the qualitative analysis of the effect that the polymer matrix has on the heat transport along the nanotube when the coupling agents are not present. By comparing the curves for BNNT and BNNT/PE, while Fig.~\ref{fig:kofom}~(a) indicates that only the lowest frequencies, about $\omega\le 10$~THz, are altered due to the presence of PE, the temp-grad method in Fig.~\ref{fig:kofom}~(b) offers a very different picture, showing a significant reduction in heat transport among a much larger frequency interval. Without the coupling agents, the interaction between PE and BNNT is only due to the long-distance van der Waals forces, which are not expected\cite{thoiut10} to affect high frequency BNNT's modes, since they are mediated by covalent bonds in the nanotube---a notion also confirmed by the strong reduction shown in Fig.~\ref{fig:kofom} at all frequencies for the composite systems with coupling agents, which  join covalently both materials. However, the two methods offer a very different picture on the extend of the frequency range in the nanotube altered by the van der Waals interaction with the polymer matrix


\section{Inconsistency of the temperature-gradient method}
\label{sec:incon}

The temp-grad method is naturally applied after the observation of a linear---time-averaged---temperature profile in the simulations. We prove in this section, however, that this linear profile actually renders the temp-grad method inconsistent when the thermal conductivity depends on the size of the material. 

\begin{figure}[h]
 \centering
 \includegraphics[width=7cm,height=4cm]{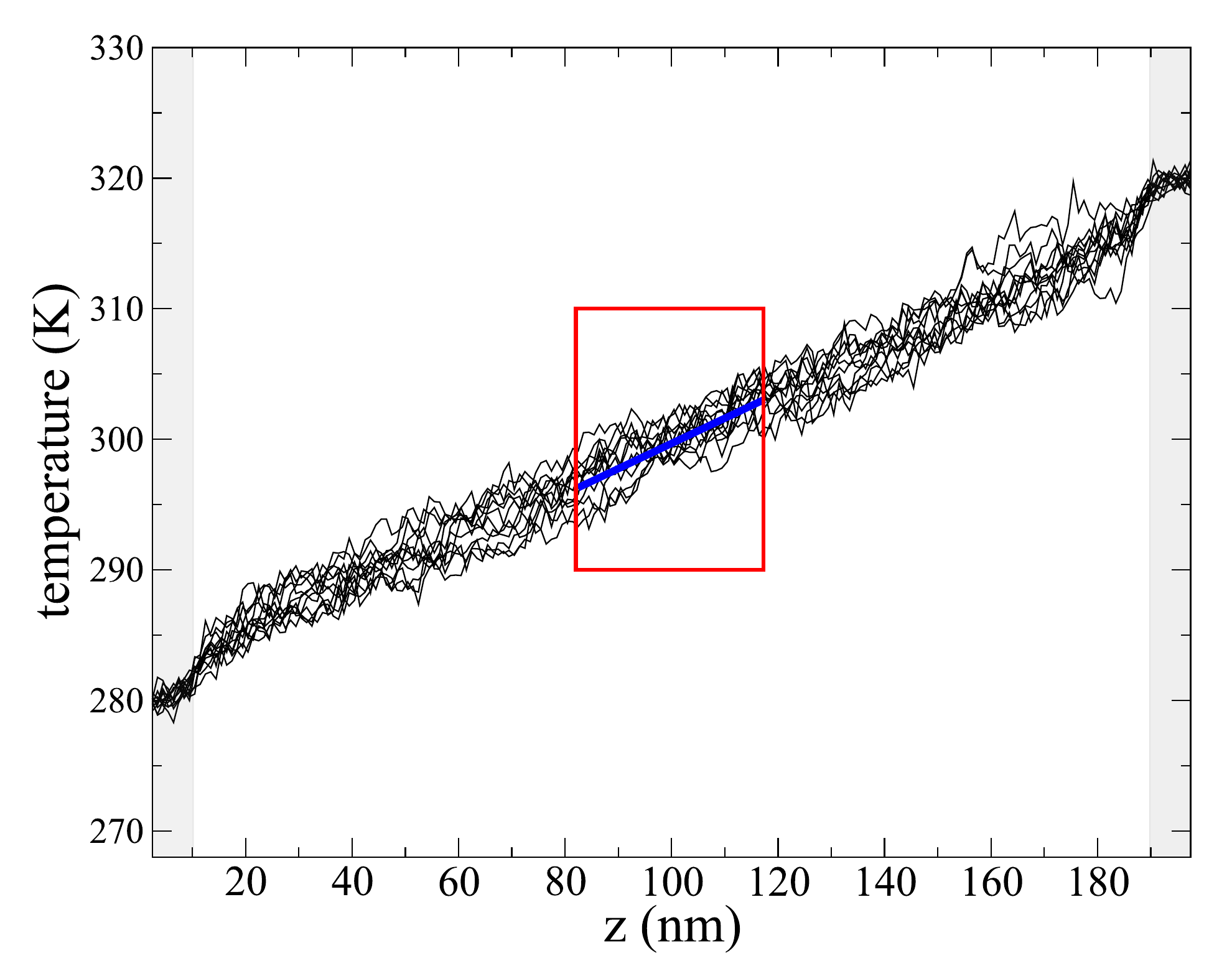}
 \caption{
Temperature profile of a simulation of bare BNNT with a non-thermostatted region of length 180~nm. A small region at the center of length $L'=35$~nm is highlighted to illustrate the inconsistency of the temp-grad approach, see main text. 
}
 \label{fig:tempgrad2}
\end{figure}

Figure~\ref{fig:tempgrad2} shows the temperature profile in a simulation of the bare nanotube with a non-thermostatted region of length 180~nm. The length of the region in the tube with linear profile is about $L=175$~nm. According to the temp-grad method, the conductivity for this length, $k_\mathrm{TG}(L)$, is obtained from the slope by using Eq.~(\ref{eq:ktg}). 

\begin{figure}[h]
 \centering
 \includegraphics[width=7cm,height=4cm]{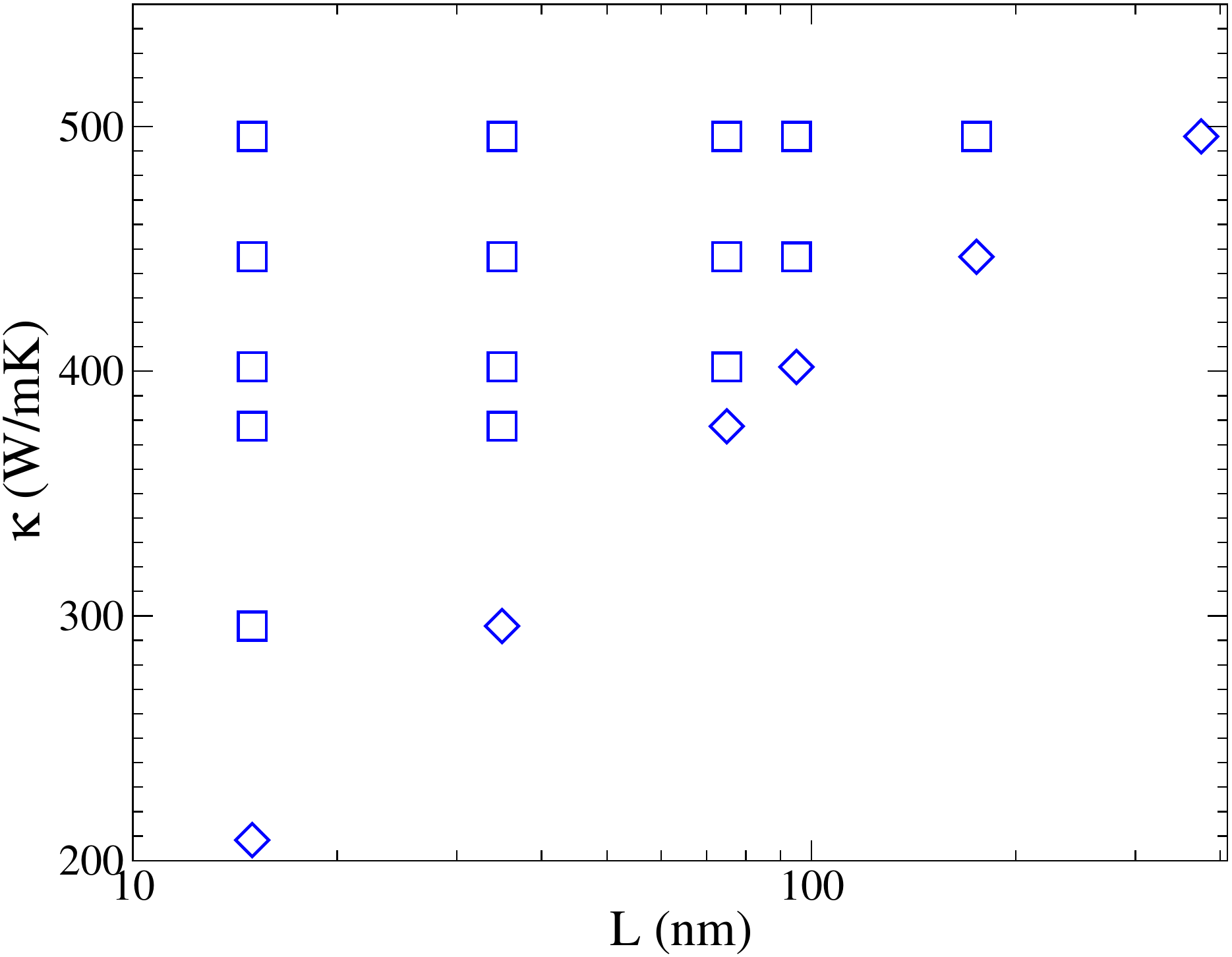}
 \caption{
Thermal conductivity of bare BNNT as a function of its length obtained with the temp-grad method with Langevin baths.  Diamonds correspond to the largest tube size that can be calculated in each simulation, and the squares to the values obtained for smaller sizes using (\ref{eq:k:tempgrad:smaller}), which exploits the local nature of the temp-grad method. The multiple values obtained for each length illustrate the inconsistency of this method for superdiffusive systems. 
}
 \label{fig:tempgrad-incon}
\end{figure}

Consider now a system consisting of a nanotube of length $L'=35$~nm.  To this effect, we can use the simulation of the larger tube shown in Fig.~\ref{fig:tempgrad2}, focus on a central region of length $L'$, and regard the other regions of the nanotube as boundaries. 
The temperature difference $T'_h-T'_c$ at these boundaries---as well as the temperature gradient in the center---is lower here than in the simulation of similar size reported in Fig.~\ref{fig:tempgrad}. This is not a problem by itself, because it just implies a smaller flux at the boundaries ---as the flux has been already verified to be directly proportional to the temperature difference or gradient.
On the other hand, we can easily compute the conductivity  $k_\mathrm{TG}(L')$ by using its definition (\ref{eq:ktg}), and the data of the larger system of length $L$ in  Fig.~\ref{fig:tempgrad2}, as
\begin{equation}\label{eq:k:tempgrad:smaller}
k_\mathrm{TG}(L')=-j' / \left(\frac{\partial T'}{\partial z}\right)_\mathrm{bulk}
=-j / \left(\frac{\partial T}{\partial z}\right)_\mathrm{bulk}=
k_\mathrm{TG}(L),
\end{equation}
because the heat flux $j$ is constant along the $z$-coordinate, and so is the slope of the temperature profile---as shown in  Fig.~\ref{fig:tempgrad2}. Therefore, according to the temp-grad method, the thermal conductivity should not depend on the system length, in direct contradiction with the numerical observations reported in Fig.~\ref{fig:kofL}, in which $k_\mathrm{TG}(L')<k_\mathrm{TG}(L)$ ---in fact about 1.5 times larger. The same calculation can be carried out for other sizes $L'$, as shown in Fig.~\ref{fig:tempgrad-incon}. This shows that the temp-grad method is an inconsistent procedure that should be avoided when the thermal conductivity depends on the size of the system. 

On the other hand, the 	above argument does not imply that a thermal conductivity with a dependency on the system size can not be well defined. The heat flux is proportional to the temperature difference at the heat baths---at least if this difference is small enough, as guaranteed by linear response theory, and confirmed by the results of Tab.~\ref{tab:k:composite}---and thus, provided it does not depend significantly on the details of heat baths used at the boundaries, as shown in Fig.~\ref{fig:kofL}, it can be meaningfully defined with the temp-diff prescription  (\ref{eq:ktd}), like it is normally done in strict one-dimensional systems. Since regions of nonlinear profile are admitted with the method, the size $L$ of the system should not be taken to that of a linear region, like in the temp-grad method, but to the full extend up to the heat baths where the temperature difference is imposed, reflecting the global nature of the thermal conductivity in superdiffusive materials.

While a certain dependence on the details of the boundaries is still expected in a superdiffusive material, it is still plausible to look for approximate expressions for $\kappa_\mathrm{TD}(L)$ which only takes into account the finite size $L$, like the standard phonon approach based on the Boltzmann equation and the relaxation time approximation \cite{mcgau04}
\begin{equation}
k_\mathrm{TD}(L)=V^{-1}\sum_n\sum_k C_{k n} v_{k n}^2\tau_{k n},
\label{eq:boltz}
\end{equation}
where $V$ is the material's volume, $C$ specific heat per mode $n$ with wave number $k$, $v$ is the phonon group velocity, and $\tau$ its the relaxation time. From (\ref{eq:boltz}), the dependence of the conductivity on the material's size is determined by the phonon mean free path $\lambda=v\tau$, as no lengths larger that $L$ are allowed. Only if the relevant mean free paths were already lower than $L$ one could find a conductivity independent of $L$---the situation at play in normal diffusive materials.

The formula (\ref{eq:boltz}) thus easily explains why the conductivity increases monotonically with the system size, as heat transporting phonons with larger mean free paths are allowed as the system becomes larger. This explanation also highlights the need to take into account the whole length of the material where the phonons can travel undisturbed from the baths' external agents, even if that includes regions with nonlinear temperature profiles.


\section{Conclusions}
\label{sec:con}

We have studied, using nonequilibrium molecular dynamics simulations, a realistic example of a superdiffusive material, a nanotube immersed in a polymer matrix. The addition of coupling agents, with various densities, allowed us to study the transition in the composite from superdiffusive to diffusive behavior. We then show that the frequently used method to compute thermal conductivities in superdiffusive nano systems, the temperature-gradient method, based on the measurement of the slope of the temperature profile, yield different, incompatible physical rationalizations of the same phenomenon when compared with a method based on the temperature difference at the heat baths at the boundaries, the latter being common in studies of strict one-dimensional systems. The simulation data also showed a better agreement in the conductivities obtained with different thermostats with the temperature-difference method. 

Furthermore, we proved that the temperature-gradient method, despite the appearance of a distinct linear temperature profile in the middle of the material, is actually inconsistent with a thermal conductivity that depends on the material size. In contrast, the definition of the conductivity based on the temperature difference finds no such difficulty, and is in better agreement with the ideas and theories based on phonon transport, such as the Boltzmann equation in the relaxation time approximation. 

Ultimately, the temperature-gradient method is flawed because it is essentially based on an assumption of a local relationship between the heat flux and the temperature gradient, while the superdiffusive nature of the material makes necessarily this relation global. However, a well-defined,  size-dependent thermal conductivity $\kappa(L)$, independent of the systems in contact with the material is also not guaranteed, and more work is needed to establish its validity.



\begin{acknowledgments}
DC acknowledges financial support from the Ministerio de Ciencia e Innovaci\'on of Spain, Grant No. PID2019-105316GB-I00. 
The computer simulations were carried out at LvLiang Cloud Computing Center of China, using TianHe-2.
\end{acknowledgments}

%

\end{document}